\documentclass[%
reprint,superscriptaddress,
amsmath,amssymb,
aps,
prb,
showpacs,
]{revtex4-1}
\usepackage{color}
\usepackage[usenames,dvipsnames,svgnames,table]{xcolor}
\usepackage{amssymb}
\usepackage{mathrsfs}
\usepackage{graphicx}
\usepackage{dcolumn}
\usepackage{bm}
\usepackage{hyperref}
\hypersetup{colorlinks=true,linkcolor=NavyBlue,citecolor=BrickRed,urlcolor=NavyBlue}

\begin{document}

\providecommand{\abs}[1]{\lvert #1 \rvert}
\providecommand{\Ham}{\mathcal{H}}
\providecommand{\bfmath}[1]{\mathbold #1 }
\newcommand\blue[1]{{\color{blue}#1}}
\newcommand\red[1]{{\color{red}#1}}
\newcommand\org[1]{{\color{orange}#1}}

\title{Hund's metal crossover and superconductivity in the 111 family of iron-based superconductors}

\author{Pablo Villar Arribi}
\affiliation{Laboratoire de Physique et Etude des Mat\'eriaux, UMR8213 CNRS/ESPCI/UPMC, Paris, France}
\affiliation{International School for Advanced Studies (SISSA), Via Bonomea 265, I-34136 Trieste, Italy}
\email{pv.arribi@gmail.com}
\author{Luca de' Medici}
\affiliation{Laboratoire de Physique et Etude des Mat\'eriaux, UMR8213 CNRS/ESPCI/UPMC, Paris, France} 
\email{demedici@espci.fr}

\date{\today}

\begin{abstract}
We study LiFeP, LiFeAs and NaFeAs in their paramagnetic metallic phase including dynamical electronic correlations within a density functional theory + slave-spin mean-field framework. The three compounds are found to lie next to the crossover between a normal and a Hund's metal, where a region of enhanced electronic compressibility that may boost superconductivity is systematically present in this type of systems. We find that LiFeP lies in the normal metallic regime, LiFeAs at the crossover, and NaFeAs is in the Hund's metal regime, which possibly explains the different experimental trends for the pressure- and doping-dependence of superconductivity in these compounds. Our picture captures the orbitally-resolved mass renormalizations measured in these materials, while an analysis of the Sommerfeld specific-heat coefficient highlights some limitations of currently used implementations of density-functional theory for the correct prediction of the details of band structures in the iron-based superconductors.
\end{abstract}
\maketitle
\section{Introduction}\label{sec:intro}

Among the different families of iron-based superconductors (IBSC), the so-called ``111'' (due to its stoichiometry), has substantially attracted less attention than the other ``122'', ``11'' and ``1111''. The origin of superconductivity in this series of compounds, and in general in all IBSC has still to be clarified. Some families (122 and 1111) exhibit long-range magnetic order, which is suppressed in favor of a superconducting phase when the compound is doped or put under pressure~\cite{Paglione_review,IBSC_springer,martinelli_phase_diagrams}. This suggests that spin-fluctuation mediated interactions could play an important role on the pairing mechanism for superconductivity~\cite{fernandes_nature_spin_fluc,Chubukov2015}. However, the general validity of this phenomenology is jeopardized in some situations. This is the case of FeSe, a stoichimetric superconductor in a nematic phase but without any long-range magnetic order~\cite{Boehmer_PRL_Tetra_Ortho}, and also the case of the 111 series, and thus other complementary explanations are needed.

LiFeAs, the parent compound of this family, is an unconventional stoichiometric superconductor below 18~K~\cite{tapp_2008_lifeas} in a tetragonal structure that displays no long-range magnetic order at low temperatures. This can be attributed to the lack of nesting of the Fermi surface in the stoichiometric compound, as seen by angle-resolved photoemision spectroscopy (ARPES) measurements~\cite{Borisenko_LiFeAs_sc_nesting_PRL.105.067002}. Besides this compound, two other isoelectronic analogues have been synthesized. If As$^{3-}$ is substituted with P$^{3-}$ one obtains LiFeP, another stoichiometric superconductor below 6~K~\cite{Deng_2009_LiFeP} also in a tetragonal structure and not magnetically ordered. On the other hand, when Li$^+$ is substituted by Na$^+$, one finds NaFeAs. This compound slightly differs from the two previous ones. At ambient temperature it is also in a tetragonal phase, but upon cooling it undergoes a structural tetragonal-to-orthorhombic transition at $T\sim54$~K into a nematic phase~\cite{Parker_NaFeAs_PRL.104.057007, Watson_NaFeAs_PRB.97.035134}. Below $T\sim45$~K, antiferromagnetic order appears and below $T\sim10$~K non-bulk superconductivity arises, coexisting with magnetism in the orthorhombic phase~\cite{Parker_2009_NaFeAs}.

 \begin{figure}[ht]
    \begin{center}
       \includegraphics[width=8cm]{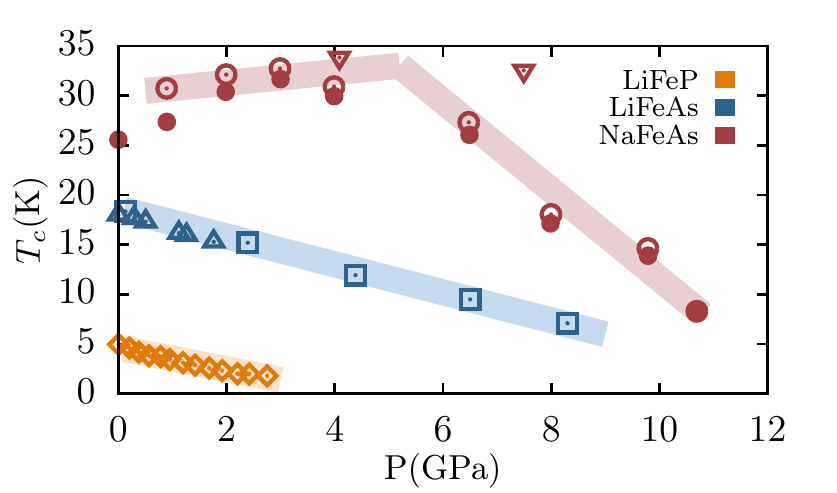}
       \caption{Summary of the pressure dependence of $T_c$ of LiFeP (yellow diamonds from Ref.~\onlinecite{Mydeen_LiFeP_pressure}), LiFeAs (blue squares from Ref.~\onlinecite{Zhang_LiFeAs_pressure} and blue triangles from Ref.~\onlinecite{Gooch_LiFeAs_pressure}), and NaFeAs (red open circles from Ref.~\onlinecite{wang_pressure_111}, red inverted triangles from Ref.~\onlinecite{Kitagawa_NaFeAs_pressure} and red solid circles corresponding to Na$_{0.86}$FeAs from Ref.~\onlinecite{Zhang_2009_NaFeAs_pressure}). Color lines are just a guide to the eye.}
       \label{fig:pressure}
    \end{center}
 \end{figure}

Charge doping and hydrostatic pressure modify superconductivity in these systems. Substitutions of Fe by Co, Ni and Cu have been achieved both in LiFeAs~\cite{Pitcher_JACS2010_NiCo, Aswartham_LiFeAs_Co_PRB, Wright_Ni_doped_LiFeAs, Xing_JPCM_LiFeAs_Cu} and in NaFeAs~\cite{Parker_NaFeAs_PRL.104.057007, Wang_NaFeAs_Cu_pressure_PRB}. They are detrimental for superconductivity in the case of LiFeAs, but for NaFeAs they destroy the magnetic order and enhance $T_c$ up to  a maximum around 5-6$\%$ in electron doping. For LiFeP, no doped samples have been synthetized so far. The effect of pressure, summarized in Fig.~\ref{fig:pressure}, is slightly similar. For LiFeP and LiFeAs, $T_c$ gets reduced for increasing pressure until it vanishes at $\sim 3$~GPa and $\sim10$~GPa respectively, being the $T_c$ of LiFeAs always higher. The scenario is again different in NaFeAs, for which $T_c$ gets enhanced at lower values of pressure (while long-range magnetic order gets destroyed~\cite{Kitagawa_NaFeAs_pressure}) up to $~33$~K at $\sim4$~GPa before getting reduced and eventually vanishing at higher values of pressure. This different behavior in such similar compounds raises a question about whether or not these materials can be understood using the same theoretical grounds and what is the physics driving such a difference.

One missing ingredient in weak-coupling theories that is key to explain some of the fundamental electronic properties in all IBSC~\cite{yin_haule_kotliar_ibsc_nature} are dynamical electronic correlations. The 111 family is no stranger in this respect. Experimentally there are several evidences that point in this direction, the main one being the presence of large effective quasiparticle masses seen in ARPES~\cite{Brouet_ARPES_PhysRevB.93.085137, He_ARPES_PhysRevLett.105.117002, Charnukha_ARPES_NaFeAs, Borisenko_LiFeAs_sc_nesting_PRL.105.067002} and Quantum Oscillations~\cite{Putzke_Coldea_PRL2012,Zeng_QO_PRB2013} measurements. These have also been predicted theoretically in calculations within dynamical mean-field theory (DMFT)~\cite{Ferber_Valenti_LiFeP_PRL.109.236403,Lee_Haule_Kotliar_111_PRL.109.177001, Ferber_Valenti_LiFeAs_PRB.85.094505,Nekrasov_NaFeAs_2015}. Moreover, these large effective masses are orbital-selective, a typical feature of IBSC~\cite{Luca_PRL_IBSC_2014}. These features together with the presence of large fluctuating paramagnetic moments (as seen for example in Co-doped NaFeAs~\cite{NaFeAs_mm_PRB.93.214506}) indicate that these materials display the phenomenology of Hund's metals~\cite{demedici_Hunds_metals}. These are systems in which the electronic properties are strongly influenced in a non-trivial way due to the large value of Hund's coupling~\cite{yin_haule_kotliar_ibsc_nature}.

Another (recently discovered) feature of Hund's metals is their proximity to a region of enhanced (and sometimes divergent) electronic compressibility that accompanies the crossover between the normal- and the Hund's-metal phase~\cite{LdM_PRL2017} that departs from the half-filled Mott transition. In the interaction-doping phase diagram this region extends into a moustache-like shape towards inconmensurate values of electronic density and into higher values of interaction strength. Such enhancement has been found both in simplified featureless models where it has been shown how the break of rotational invariance extends it to larger values of doping~\cite{Mary_PRB2020}, and also in realistic simulations of different IBSC~\cite{LdM_PRL2017,Edelmann_Chromium_analogs_PRB2017, PVA_LdM_PRL2018}. This suggests that it could potentially act as a superconductivity booster in Hund's metals~\cite{LdM_PRL2017} since it can be related to enhanced quasiparticle interactions and Fermi liquid instabilities like superconductivity~\cite{EmeryKivelson, GrilliRaimondi_IntJModB, CDG_PRL95}. Moreover, a two-dimensional many-variable Variational Monte Carlo (mVMC) study in LaFeAsO has found a zone of electronic phase separation~\cite{Misawa_LaFeAsO} in a similar region of parameters that can be related to the same electronic instabilities. 

In this study, we address the situation for the 111 family of IBSC by characterizing this crossover and the electronic compressibility of these materials, and we propose a unified picture that can explain the different experimental trends both for superconductivity and some of the normal-state electronic transport properties. The article is organized as follows: in sec.~\ref{sec:models} we discuss the model and the chosen method we use to solve it. In sec.~\ref{sec:results} we present the results. First we comment on the trends in the electronic correlations and in the electronic compressibility (sec.~\ref{sec:correlations}), and we follow with the results for the mass renormalizations and Sommerfeld coefficient (sec.~\ref{sec:sommerfeld}). Finally, in sec.~\ref{sec:conclusions} we summarize the key aspects of this work.

\section{Model and methods}\label{sec:models}

All the members of the 111 family of IBSC have five bands of mainly Fe-$3d$ character near the Fermi level. To model them we use a 5-orbital Hubbard-Kanamori Hamiltonian $\hat{\mathcal{H}}-\mu\hat{N}=\hat{\mathcal{H}}_0+\hat{\mathcal{H}}_{int}-\mu\hat{N}$. This includes a non-interacting term $\hat{\mathcal{H}}_0$ and an interacting term to account for dynamical electronic correlations present in these systems, $\hat{\mathcal{H}}_{int}$. $\mu$ is the chemical potential and $\hat{N}$ the total number of particles. 

The non-interacting part can be written as:
\begin{equation}
\hat{\mathcal{H}}_0=\sum_{i\neq j,m,m',\sigma}t^{mm'}_{ij}d^{\dagger}_{im\sigma}d_{jm'\sigma}+\sum_{i,m,\sigma}\varepsilon_m\hat{n}_{im\sigma},
\label{eq:H0}
\end{equation}
where $d^{\dagger}_{im\sigma}$ creates an electron with spin $\sigma$ in orbital $m=1,...,5$ on the site $i$ of the lattice, and $\hat{n}_{im\sigma}=d^{\dagger}_{im\sigma}d_{im\sigma}$ is the number operator. The hopping integrals $t^{mm'}_{ij}$ and on-site orbital energies $\varepsilon_m$ are calculated in a basis of maximally-localized Wannier functions~\cite{Wannier_review, Wannier90}. These correspond to a tight-binding parametrization of the bare band structure, which is calculated within the density-functional theory (DFT) framework as implemented in the code {\sc Wien2k}~\cite{WIEN2k}. For all the calculations we use the GGA-PBE exchange-correlation functional~\cite{PBE}, and the experimental lattice parameters and atomic positions from Refs.~\onlinecite{Deng_2009_LiFeP, tapp_2008_lifeas, Parker_2009_NaFeAs}.

The term including the many-body electron-electron interactions reads:
\begin{equation}
\label{eq:Hint}
\begin{split}
\hat{\mathcal{H}}_{int}&=U\sum_{m}\hat{n}_{m\uparrow}\hat{n}_{m\downarrow}+(U-2J)\sum_{m\neq m'}\hat{n}_{m\uparrow}\hat{n}_{m'\downarrow}\\
&+(U-3J)\sum_{m<m',\sigma}\hat{n}_{m\sigma}\hat{n}_{m'\sigma},
\end{split}
\end{equation}
where $U$ is the local on-site intra-orbital Coulomb repulsion, and $J$ the Hund's coupling. For these compounds we choose $U=3.2$ eV, as obtained for LiFeAs by ab-initio constrained random-phase approximation calculations~\cite{Miyake} (cRPA), and we assume it is the same for LiFeP and NaFeAs (although we will perform scans in $U$). We fix $J/U=0.225$ for all our calculations, according to the prescription described in section S2 from Ref.~\onlinecite{PVA_LdM_PRL2018}.

We solve this many-body problem using slave-spins mean-field theory (SSMFT)~\cite{luca_massimo_ssmft}. This is a very convenient approach to study IBSC since it describes by construction a Fermi liquid (which is the low-temperature behaviour displayed by these materials in their normal phase~\cite{RullierAlbenque_Fermi_liquid}), it correctly captures the orbital differentiation of these compounds~\cite{Luca_PRL_IBSC_2014}, and it accurately predicts the Sommerfeld coefficient of the 122 family~\cite{hardy_luca_122}. 

SSMFT yields an effective quasiparticle Hamiltonian~\cite{luca_massimo_ssmft} of the form:
\begin{equation}
\hat{\mathcal{H}}_{QP}=\!\!\!\!\!\!\!\!\sum_{i\neq j,m,m',\sigma}\!\!\!\!\!\!\!\!\sqrt{Z_mZ_{m'}}t^{mm'}_{ij}f^{\dagger}_{im\sigma}f_{jm'\sigma}+
\sum_{i,m,\sigma}(\varepsilon_m-\lambda_m)\hat{n}^f_{im\sigma},
\label{eq:Heff}
\end{equation}
where $f^{\dagger}_{im\sigma}$ 
creates a quasiparticle with corresponding quantum numbers. The renormalization due to the interactions in eq.~(\ref{eq:Hint}) in SSMFT is brought in by the factors $Z_m$ (that act as the inverse of the mass enhancement) and $\lambda_m$ (that shift the on-site energy)~\footnote{In order for $\lambda_m$ to vanish at zero interactions a shift $\lambda^0_m$ is necessary, as analyzed in Appendix A of Ref.~\onlinecite{Mary_PRB2020}. We use here the prescription $\lambda^0_m=\frac{(2n_{m\sigma}-1)}{n_{m\sigma}(1-n_{m\sigma})}h_{m\sigma}$ evaluated self-consistently at all interactions. We have performed sample checks with the alternative prescription from Ref.~\onlinecite{yu_si_prb2012} obtaining almost identical results which confirm the robustness of our calculations with respect to this technical aspect of the method.}. These factors are calculated in a set of self-consistent mean-field equations that involve the auxiliary slave-spin variables~\cite{luca_massimo_ssmft}, and depend on all the physical parameters of the problem. The electronic compressibility is simply $\kappa_{el}=\frac{d n}{d \mu}$, where $n$ is the charge density which following Landau's Fermi liquid theory~\cite{nozieres} must be equal to the total number of quasiparticles $n_f$:
\begin{equation}
n_f\equiv \sum_{km\sigma}\langle f^\dagger_{km\sigma}f_{km\sigma} \rangle=\int^\mu d\varepsilon D^*(\varepsilon)=n,
\label{eq:density}
\end{equation}
where $D^*(\varepsilon)$ is the renormalized (quasiparticle) density of states (DOS).

\section{Results and discussion}\label{sec:results}
\subsection{Correlation strengths and electronic compressibility}\label{sec:correlations}

 \begin{figure}[ht!]
    \begin{center}
       \includegraphics[width=8.5cm]{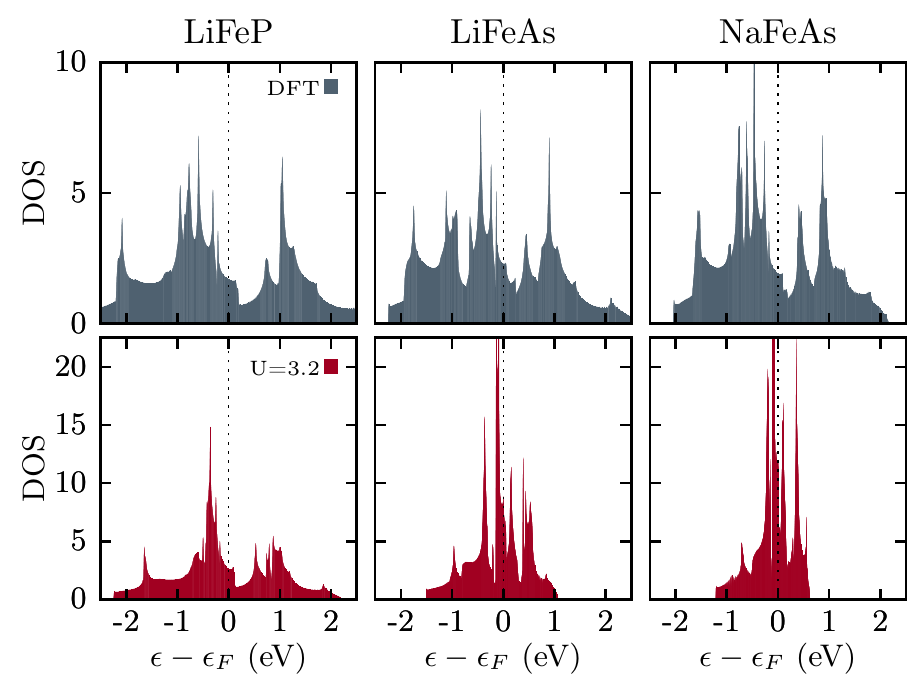}
       \caption{Densities of states of the electronic band structure for LiFeP, LiFeAs and NaFeAs calculated with DFT (upper panels) and with DFT+SSMFT for a value of the local Coulomb interaction of $U=3.2$~eV and $J/U=0.225$ (lower panels).}
       \label{fig:DOS}
    \end{center}
 \end{figure} 
 
In Fig.~\ref{fig:DOS} we show a comparison between the DOS of LiFeP, LiFeAs and NaFeAs, calculated at the DFT level ($U=0$~eV) and at the relevant value of the interaction for these materials ($U=3.2$~eV) with DFT+SSMFT. The total bare bandwidth of the isolated Fe-$3d$ manifold is $\sim5.6$~eV for LiFeP, $\sim4.9$~eV for LiFeAs and $\sim4.3$~eV for NaFeAs. This is expected simply by looking at the difference in the lattice parameters in the three compounds. A larger overlap between Fe-$3d$ orbitals is expected for LiFeP, which has the smallest lattice spacing, followed by LiFeAs and finally by NaFeAs, in which the Fe atoms are the furthest from one another, thus having the lowest kinetic energy of the three compounds. If we simply consider the degree of correlation as the direct competition between bare kinetic energy and on-site Coulomb repulsion, this difference in bandwidths directly implies that there will be a hierarchy in the degree of correlation for these materials once local interactions are taken into account. This is indeed the result we obtain in our calculations. For the same values of the interaction parameters $U$ and $J$, the overall degree of correlation (as seen by the dispersion of the renormalized quasiparticle dispersion) is larger for NaFeAs (which has the narrowest of the renormalized densities of states), followed by LiFeAs and finally by LiFeP as showed in Fig.~\ref{fig:DOS}. This has been confirmed by transport measurements~\cite{kasahara_2012_transport}, which indicate that LiFeP is less correlated than LiFeAs.

 \begin{figure}[ht]
    \begin{center}
       \includegraphics[width=8.5cm]{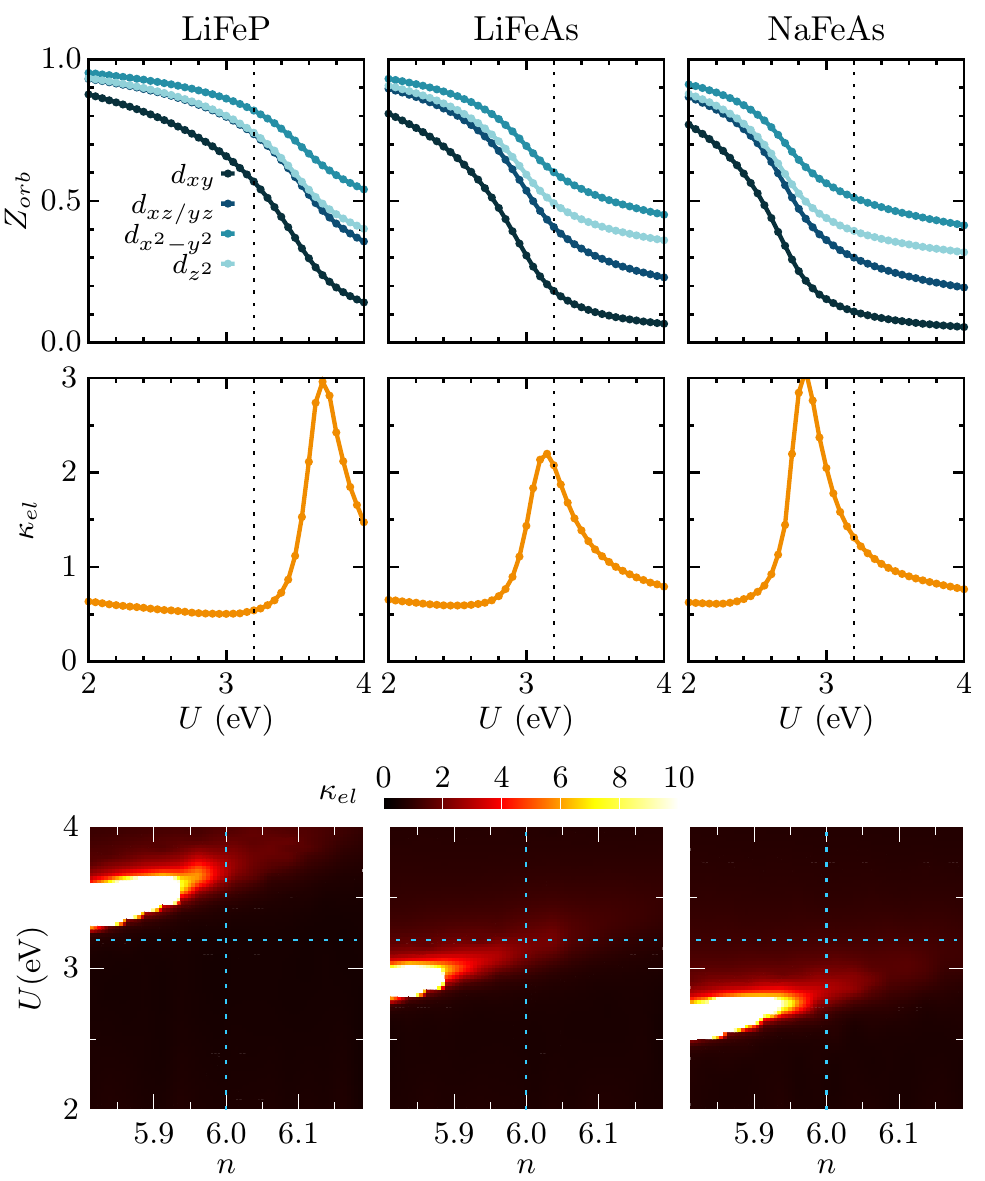}
       \caption{Upper panels: orbitally resolved quasiparticle weights $Z_{m}$ as a function of the local Coulomb repulsion $U$ and for $J/U=0.225$ for LiFeP, LiFeAs and NaFeAs.  Mid panels: electronic compressibility for the same compounds also as a function of $U$. Lower panels: heat map of the electronic compressibility as a function of $U$ and the electronic density $n$. The dotted lines represent the position of the stoichimetric compounds in these phase diagrams, this is, at $U=3.2$ eV and $n=6.0$.}
       \label{fig:Z_vs_U_111}
    \end{center}
 \end{figure}

According to the general phenomenology of Hund's metals~\cite{demedici_Hunds_metals}, as the interaction strength gets larger, not only the overall degree of correlation will increase, but also the orbitally-resolved renormalization factors will become progressively more differenciated. We thus expect a difference in the orbitally-resolved quasiparticle weights $Z_m$ among these materials. This can be observed tracking the evolution of these factors as a function of the local Coulomb interaction strength $U$ (showed in the upper panels in Fig.~\ref{fig:Z_vs_U_111}). For the same value of $U$, LiFeP shows a smaller dispersion in the values of $Z_m$ than LiFeAs (which is slighly more correlated) and than NaFeAs (the most correlated of the three compounds according to our simulations). This trend has been comfirmed experimentally looking at the measured orbitally resolved mass enhancements (which in our theory are just $Z_m^{-1}$) from Quantum Oscillations and sheet-resolved ARPES experiments~\cite{Putzke_Coldea_PRL2012, Putzke_Coldea_PRL2012, Zeng_QO_PRB2013, Brouet_ARPES_PhysRevB.93.085137, Borisenko_LiFeAs_sc_nesting_PRL.105.067002, He_ARPES_PhysRevLett.105.117002}. We will discuss this point more in depth in the following section.

We turn now our attention to the electronic compressibility $\kappa_{el}$. The results of our simulations are shown in the mid panels of Fig.~\ref{fig:Z_vs_U_111}, where we track the evolution of $\kappa_{el}$ on the different members of the 111 family as a function of $U$. In all the three materials, an enhancement appears around the same values of $U$ where the decrease in the orbitally-resolved quasiparticle weights is more pronounced (upper panels in Fig.~\ref{fig:Z_vs_U_111}). For LiFeP the peak in $\kappa_{el}$ is located at a larger value of $U=3.7$~eV and for LiFeAs it occurs at $U=3.15$~eV. For NaFeAs, the peak appears at $U=2.8$~eV, slightly below the relevant value of $U$ for these materials. Such enhancement has been shown to occur at the crossover between a normal and a Hund's metal and linked to supercondictivity~\cite{LdM_PRL2017}.  Indeed, in previous works, we have found for realistic simulations in other IBSC~\cite{LdM_PRL2017, PVA_LdM_PRL2018} a link between this enhancement in $\kappa_{el}$ and the experimental trends of superconductivity. In particular, for FeSe it has been shown~\cite{PVA_LdM_PRL2018} how increasing hydrostatic pressure can effectively move that peak in $\kappa_{el}$ closer to the relevant values of $U$ for that material, correlating positively with the experimentally-observed trends for superconductivity in FeSe under pressure. Furthermore, for FeSe monolayer, this enhancement is not only present in the system, but the region of instability is much larger, which can be explained due to the difference in the crystal-field splitting of the bare hamiltonians of FeSe bulk and FeSe monolayer~\cite{Mary_PRB2020}. Also the value of $\kappa_{el}$ is higher, which also correlates with the higher value of $T_c$ seen experimentaly in the monolayer.

If we look at the experiments for the 111 family under pressure in Fig.~\ref{fig:pressure} we see that for LiFeP and LiFeAs, increasing hydrostatic pressure directly suppresses the $T_c$. This is compatible with our picture, since pressure effectively decorrelates the system, thus moving the peak in $\kappa_{el}$ to larger values of $U$. Hypotetically, we can also explain why $T_c$ in stoichiometric LiFeAs is higher than in LiFeP, since the former is located closer to that region of enhanced $\kappa_{el}$. In the case of NaFeAs, increasing pressure enhances superconductivity. The experiments show a maximum in $T_c$ for this compound around $\sim$4~GPa~\cite{wang_pressure_111, Kitagawa_NaFeAs_pressure}. This behavior is also supported our picture, because the peak in $\kappa_{el}$ is now located at a lower value of $U$ than that for the compound. Incresing pressure will move this peak to higher values of interaction, closer to $U\sim3.2$ (as it happens for FeSe~\cite{PVA_LdM_PRL2018}) and eventually will cross that value after which the peak moves away from it. All this implies that similarly to how hydrostatic pressure effectively moves this crossover at higher values of $U$, isoelectronic chemical substitution can be used in a similar fashion to tune the crossover between a normal and a Hund's metal, with the advantage that it can act both as effective positive or negative pressure. Introducing different isovalent ions either in the spacing layer (Li$^{+}\rightarrow$Na$^{+}$) or in the Fe-ligand layers (P$^{3-}\rightarrow$As$^{3-}$) modifies the Fe-Fe distance, thus effectively displacing the crossover.

In order to further validate our scenario, we have also calculated the electronic compressibility for these materials at different doping levels, that we illustrate in the interaction-doping ($U$ vs. $n$) phase diagrams shown in the lower panels in Fig.~\ref{fig:Z_vs_U_111}. We observe a moustache-like region of enhanced $\kappa_{el}$ that is present for all the three compounds. Such a region extends through different values of $U$ and $n$ for each material, which means that, according to our picture, $T_c$ in these systems would be affected differently by doping (with electrons or holes) and pressure.  

For LiFeAs, the location of the crossover in the $U$ vs. $n$ parameter space (as indicated by the region of enhanced $\kappa_{el}$) implies that electron doping progressively reduces the value of $\kappa_{el}$, and thus $T_c$ should decrease accordingly. This is confirmed experimentaly, as superconductivity is rapidly suppressed by electron doping substituting Fe with Co, Ni or Cu~\cite{Pitcher_JACS2010_NiCo, Aswartham_LiFeAs_Co_PRB, Wright_Ni_doped_LiFeAs, Xing_JPCM_LiFeAs_Cu}. These substitutions, in low concentrations, are though to just modify the charge density of the system in several IBSC~\cite{Ideta_PRL110.107007,McLeod_2012}, thus confirming the robustness of our scenario.

NaFeAs, according to our calculations, is well inside the Hund's metal regime. Experimentally, Co-substitution of Fe increases $T_c$~\cite{Parker_NaFeAs_PRL.104.057007} up to $\sim2.5\%$ doping before starting to suppress it at larger densities, but this increase can be adscribed to the presence of magnetism, which is suppressed by electron-doping. However, there is clear evidence that in NaFe$_{1-x}$Co$_x$As~\cite{Wang_2012_NiFeCoAs_pressure} and NaFe$_{1-x}$Cu$_x$As~\cite{Wang_NaFeAs_Cu_pressure_PRB} superconductivity is enhanced by hydrostatic pressure at sufficiently large values of electron doping where no magnetism is present in the sample. Moreover, this enhancement is seen for several values of doping, which is consistent with the position and shape of the moustache for this compound. 


\subsection{Mass renormalizations and Sommerfeld coefficient}\label{sec:sommerfeld}
 \begin{figure}[ht]
    \begin{center}
       \includegraphics[width=8.5cm]{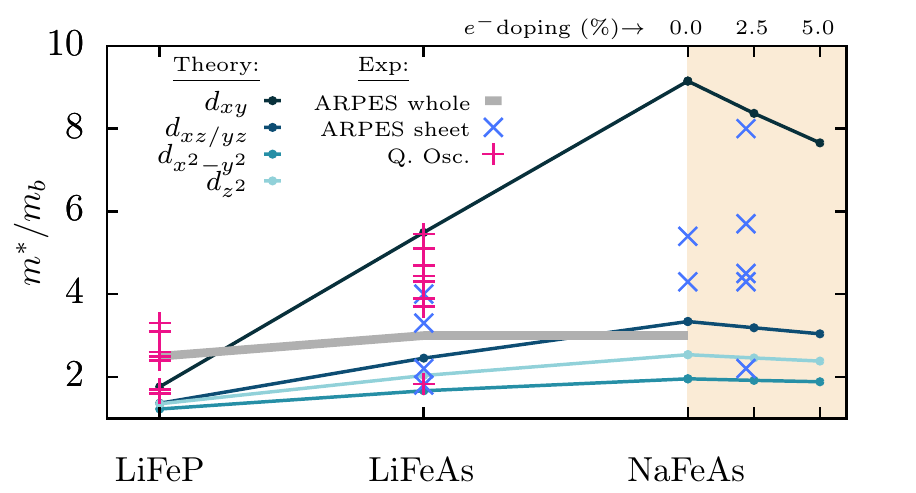}
       \caption{Orbitally-resolved mass enhancements calculated with DFT+SSMFT for a intra-orbital Coulomb interaction $U=3.2$ eV and $J/U=0.225$ and mass enhancements from Quantum Oscillations, single and sheet-resolved ARPES measurements for LiFeP~\cite{Putzke_Coldea_PRL2012}, LiFeAs~\cite{Putzke_Coldea_PRL2012, Zeng_QO_PRB2013, Brouet_ARPES_PhysRevB.93.085137, Borisenko_LiFeAs_sc_nesting_PRL.105.067002} and NaFeAs (both stoichiometric~\cite{He_ARPES_PhysRevLett.105.117002} and for $2.2\%$ Co substitution~\cite{Charnukha_ARPES_NaFeAs}).}
       \label{fig:Mass_renorm}
    \end{center}
 \end{figure}

In Fig.~\ref{fig:Mass_renorm} we show a collection of experimental electronic effective masses for LiFeP, LiFeAs and NaFeAs, obtained by ARPES and Quantum Oscilation measurements in comparison with our estimates from DFT+SSMFT calculations. Both experimental and theoretical mass renormalizations are larger and more orbital-selective going from LiFeP to LiFeAs and to NaFeAs. Our predictions also signal that the $d_{xy}$ orbital is the more renormalized of all, in agreement to what is seen by many experiments~\cite{Putzke_Coldea_PRL2012, Zeng_QO_PRB2013, Brouet_ARPES_PhysRevB.93.085137, He_ARPES_PhysRevLett.105.117002, Charnukha_ARPES_NaFeAs} and with previous theoretical calculations both with DMFT~\cite{Ferber_Valenti_LiFeP_PRL.109.236403, Lee_Haule_Kotliar_111_PRL.109.177001, Ferber_Valenti_LiFeAs_PRB.85.094505, Nekrasov_NaFeAs_2015} and with SSMFT~\cite{qimiao_lifeas_2021}.

According to our picture, if a material is in the Hund's metal regime or at the crossover with a normal metallic phase, electron doping should rapidly decrease the degree of correlation. This has been seen by ARPES as an increase of the Fermi velocities in LiFeAs and NaFeAs~\cite{Ye_ARPES_PRX2014}. For NaFeAs, however, the evolution of the mass enhancements with electron doping (shaded area in Fig.~\ref{fig:Mass_renorm}) seems to go in the opposite direction of what out theory predicts. These estimates are highly sensitive to the fitting procedure of the ARPES spectra to the DFT band structure. For instance, comparing the dispersion of the band with mainly $d_{xy}$ character from several ARPES spectra~\cite{Ge_PRX3.011020, Watson_NaFeAs_PRB.97.035134, He_ARPES_PhysRevLett.105.117002, Charnukha_ARPES_NaFeAs} of NaFeAs at similar values of doping and temperature, and in the same crystallographic phase, one can observe that the slope around the Fermi level does not decrease as a function of electron doping, as it will need to happen for an effective mass enhancement. Indeed, the opposite has been reported~\cite{Ye_ARPES_PRX2014} by looking at the bandwith of the so-called $\beta$-band (of mainly $d_{xz/yz}$ character) which increases as a function of electron doping. This will correspond to a decrease in the mass enhancement with doping, which is again consistent with our picture.

 \begin{figure}[ht]
    \begin{center}
       \includegraphics[width=8.5cm]{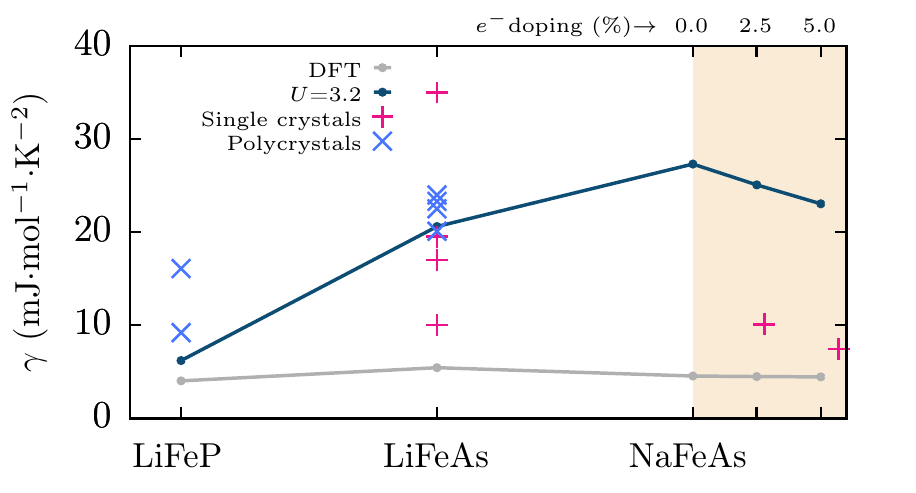}
       \caption{Collection of experimental value of the Sommerfeld coefficient $\gamma$ measured in the 111 family both in polycrystalline samples~\cite{Deng_2009_LiFeP, Kim_LiFeP_PhysRevB.87.054504, Singh_LiFeAs_somm, Chu_111_physC, Baker_2009_lifeas} and in single crystals~\cite{Stockert_LiFeAs_PRB, Wei_LiFeAs_PhysRevB.81.134527, Lee_2010_LiFeAs, Jang_PhysRevB.85.180505, Wang_NaFeAs_PhysRevB.85.224521, Chen_NaFeAs_PRL, Wang_NaFeAs_Cu_pressure_PRB}, compared to our calculations both with (U=3.2eV, J/U=0.225) and without (simply labeled "DFT") interactions.}
       \label{fig:Sommerfeld}
    \end{center}
 \end{figure} 
As seen in Fig.~\ref{fig:DOS}, the squeezing the Fe-$3d$ bands due to electronic correlations causes an enhancement in the quasiparticle DOS around the Fermi level. The Sommerfeld coefficient $\gamma$ (the low-temperature linear coefficient of the specific heat) can be directly calculated from these quasiparticle DOS and compared with the available experimental data. Our calculations show a considerable enhancement for the three compounds in comparison with the values predicted by DFT, as shown  in Fig.~\ref{fig:Sommerfeld}. The DFT calcuations show a rather low and constant value of $\gamma\sim4-5$ in comparison with the experimental results, where much larger values of $\gamma$ are observed for all the compounds. The results with DFT+SSMFT are in better agreement with the experiments. 

For LiFeP, the experimental values are around 9-16~mJ$\cdot$mol$^{-1}\cdot$K$^{-2}$. The theoretical predictions for this compound are 4.0 and 6.2$\cdot$mol$^{-1}\cdot$K$^{-2}$ with DFT and DFT+SSMFT respectively. The inclusion of correlations improves the calculated value, yet it is below the experimental one. This could be improved if we could use the values of $U$ and $J$ calculated with cRPA directly for LiFeP (unavailable at the moment in the literature), instead of using those for LiFeAs. 

In the case of LiFeAs, the available experimental data differs quite a lot, in particular in the case of single crystals, displaying values of $\gamma$ on the range 10-35~mJ$\cdot$mol$^{-1}\cdot$K$^{-2}$ with two samples displaying values $\sim20$~mJ$\cdot$mol$^{-1}\cdot$K$^{-2}$, similar to that of most of the polycrystalline samples. The quality of the samples in the beginning of the iron-pnictide era was still to be optimized, in particular for single crystals. This can be the cause of the big discrepancies seen in the experimental data. Also, all of these systems are extremely air sensitive due to its content in alcaline ions and tend to oxidize quickly. This could be another potential source of discrepancy. Focusing in the more clustered data we see that the prediction from DFT+SSMFT gives a value $\sim3.5$ times larger compared to that of DFT, matching most of these experimental measurements, in particular those in polycrystalline samples.

The case of NaFeAs is more complex. According to our calculations, this is the most correlated of the three compounds. However, because it develops antiferromagnetic order below ~60~K, the Sommerfeld coefficient has to be obtained from a very sensitive extrapolation of the data in the high-temperature paramagnetic phase. NaFeAs can be electron-doped with either Co or Cu~\cite{Parker_NaFeAs_PRL.104.057007, Wang_NaFeAs_Cu_pressure_PRB}, which suppresses the magnetic phase. This allows to access the value of $\gamma$ from an extrapolation at progressively lower temperatures, and thus more accurately. In principle, a decrease of correlations should imply a reduction of $\gamma$ (although this relation can be non-trivial and depends on the particular details of the electronic band structure). This trend is seen experimentally (shaded area in Fig.~\ref{fig:Sommerfeld}) and is correctly confirmed by our calculations. However, the value we obtain for $\gamma$ is much higher. A plausible explanation for this mismatch is the limitation of standard GGA-DFT - that we use here - in quantitatively describing the Fermi surface of IBSC. It is well known indeed that the experimentally measured size of the Fermi pockets it typically smaller than the prediction of LDA or GGA, and this is hardly cured by the inclusion of local dynamical correlations with e.g. DMFT\cite{yin_haule_kotliar_ibsc_nature} or SSMF. In fact either dynamical\cite{Ortenzi:2009,Fanfarillo:2016} or statical\cite{Jiang:2016,Scherer:2017} k-dependent corrections to the electronic self-energy are needed to compensate this mismatch, which can culminate in the complete sinking of a band predicted by standard DFT to be cutting the Fermi level, and of the disappearance of the corresponding Fermi pocket. In an ab-initio framework this typically requires the use of more computationally expensive methods - like e.g. quasiparticle GW~\cite{Tomczak_PRL2012, Tomczak_2015}- that treat more accurately than LDA/GGA the non-local electronic interactions, and in particular the screened Fock exchange\cite{vanRoekeghem:2014}. Among them, a recently proposed extension of the DFT+SSMFT approach using hybrid functionals at the DFT level~\cite{gorni2021_FeSe} has been proven to improve the description of FeSe fermiology. There the sinking of the Fermi pocket of main $d_{xy}$ character in the center of the Brillouin zone (missed by LDA/GGA) leads to correctly predicting several transport properties that are extremely sensitive to these features of the quasiparticle dispersion near the Fermi level, and the Sommerfeld coefficient. 
In the same spirit, it has been recently shown that ARPES data for LiFeAs can be interpreted as coming by a dynamical, local self-energy, once the reference non-interacting Hamiltonian is calculated with quasiparticle GW and thus beyond standard DFT~\cite{kim2020_lifeas}. 

For NaFeAs, ARPES experiments suggest that the band with $d_{xz/yz}$ character around the $\Gamma$ point is located below the Fermi energy~\cite{Ge_PRX3.011020, Watson_NaFeAs_PRB.97.035134, He_ARPES_PhysRevLett.105.117002, Charnukha_ARPES_NaFeAs}, while within our DFT(GGA)+SSMF method, this band appears just above it. We thus prospect that one of the aforementioned improvement on the standard DFT treatment could make this band sink below the Fermi level, with the corresponding suppression in spectral weight that would reduce the value of $\gamma$, in a similar fashion as it happens for FeSe~\cite{gorni2021_FeSe}. These corrections, that should improve our calculated value of the Sommerfeld coefficient, will however barely affect the electronic compressibility or the quasiparticle renormalization factors that we capture correctly, that depend on all the energy scales of the system and are much less sensitive to small changes in the quasiparticle dispersion around the Fermi level. The use of these more computationally expensive techniques in our calculations is beyond the scope of this work and is left for future investigation.
 
\section{Summary}\label{sec:conclusions}
In summary we have studied the electronic compressibility of the 111 family of IBSC with a DFT+SSMFT scheme. We have shown that all the compounds are in close proximity of a region of enhanced electronic compressibility that may be relevant for superconductivity. While LiFeAs lies at the crossover between a normal and a Hund's metal, LiFeP is clearly below that crossover, being less correlated and in the normal metallic phase, and NaFeAs is in the Hund's metal regime displaying large and orbital-selective quasiparticle effective masses. We can successfully explain the experimental trends seen for $T_c$ in these compounds under pressure and with doping, finding another confirmation of the conection between superconductivity and an enhanced electronic compressibility, thus further validating previous results in other IBSC~\cite{LdM_PRL2017, PVA_LdM_PRL2018}. Moreover, we have shown how isoelectronic chemical substitution can be effectively used to modify the proximity of IBSC to this region of instabilities in these materials. Also, our simulations correctly predict the effective mass enhancements seen experimentally in the 111 family of IBSC, improve the description of the Sommerfeld coefficient compared to the DFT estimates, and provide more insight regarding the experimental observations in specific heat measurements for these materials. We have also confirmed that a DFT+SSMFT scheme is sufficient to describe the basic electronic properties in the 111 family of IBSC in their normal paramagnetic phase. The few remaining discrepancies, for instance in NaFeAs, are probably due to the reference bare DFT hamiltonian, which has been recently shown that can be improved~\cite{gorni2021_FeSe,kim2020_lifeas}, in particular obtaining a better description of the fine details of the band structure around the Fermi level, while maintaining essentially unchanged the fundamental local many-body physics we highlight in this work.

\acknowledgments
The authors wish to thank T. Gorni for insightful discussions concerning the Sommerfeld coefficient calculations, and F. Hardy for his help assessing the heat-capacity measurements. The authors are financially supported by the European Commission through the ERC-StG2016, StrongCoPhy4Energy, GA No724177.

%

\end{document}